\documentclass[usenatbib,twocolumn,useAMS]{mn2e}

\usepackage{amsfonts}
\usepackage{amsmath}
\usepackage{graphicx}
\usepackage{color}

\newcommand{\mnras}{MNRAS}

\newcommand{\aj}{AJ}
\newcommand{\apj}{ApJ}
\newcommand{\apjl}{\apj}
\newcommand{\apjs}{\apj Suppl.}

\title[MgII absorbers and their neighbours]
       {MgII absorption systems and their neighbouring galaxies from 
        a background subtraction technique}

\author[M. A. Caler, R. K. Sheth \& B. Jain]
       {Michelle A. Caler, Ravi K. Sheth \& Bhuvnesh Jain
        \thanks{E-mail: mcaler, shethrk, bjain@physics.upenn.edu}\\
       Department of Physics and Astronomy, 
       University of Pennsylvania, 209 S. 33rd Street, 
       Philadelphia, PA 19104}

\begin{document}
\pagerange{\pageref{firstpage}--\pageref{lastpage}}

\maketitle

\label{firstpage}

\begin{abstract}
We estimate the absolute magnitude distribution of galaxies which lie within a few hundred kiloparsecs of Mg II absorption systems.  The absorption systems themselves lie along 1880 lines of sight to Sloan Digital Sky Survey Data Release 3 QSOs, have rest equivalent widths greater than 0.88\AA, and redshifts that range from $0.37 \leq z \leq 0.82$. Our measurement is based on all galaxies identified by the SDSS photometric pipeline which lie within a projected distance of about 900~$h^{-1}$ kpc of each QSO demonstrating absorption.  The redshifts of these projected neighbors are not available, so we use a background subtraction technique to estimate the absolute magnitude distribution of true neighbors.  Our method exploits the fact that, although we do not know the redshifts of the neighbors, we {\em do} know the redshift of the absorbers.  The absolute magnitude distribution we find is well described by a bell-shaped curve peaking at rest-frame $M_{\rm B} \approx -20.2$, corresponding to $L/L_* \approx 1.4$.  A comparison of this observed distribution to expected shapes derived from the literature suggests that it is unlikely to be drawn from a population that is dominated by late-type galaxies. However, when the sample is divided in half on the basis of rest-frame equivalent width, we find that some stronger equivalent width systems appear to be associated with later galaxy types.  We use these shapes, along with the observed covering fraction of $\approx 8\%$, to estimate the extent of the MgII distribution around a galaxy. For an $L_*$ galaxy, this scale is about 70$h^{-1}$~kpc.  We provide an analytic description of the method, which helps build intuition, and show that it is generally applicable to any dataset in which redshifts are only available for a small sub-sample.  Hence, we expect it to aid in the analysis of galaxy scaling relations from photometric redshift datasets.
\end{abstract}



\section{Introduction}
QSO absorption line systems have been the subject of numerous studies since their discovery and identification in the late 1960s \citep{bac68,bls68,bas69}.  Historically, these systems have been identified in spectra taken from the ground; at high enough redshift, atomic transitions with lines in the UV are redshifted into the atmospheric optical window, and indeed many such spectral lines have been used to identify these systems.  Detailed studies of the number and kinematics of these lines have greatly aided our understanding of the physical environment of the gas in these systems \citep{cv01} and have provided constraints on the amount of neutral gas in the universe at high redshift \citep{pro05}.  The $\lambda2796\lambda2803$ doublet of singly ionized magnesium (Mg II) is a popular target of spectroscopic searches due to its relative ease of identification in spectra and its association with neutral hydrogen \citep{rt00}.  This makes it an excellent probe of neutral gas, particularly at redshifts below which Lyman alpha absorption is still outside the window observable from the ground.

The connection between Mg II absorption systems and luminous galaxies has been well established \citep{bab91}, and models which place the absorbing gas in the haloes of such galaxies have had some success in explaining the absorption characteristics seen \citep{mam96,steidel02,laz}. However, a more detailed connection between the absorption systems and galaxy morphology as well as that between the absorption systems and location within the galaxies is still uncertain \citep[but see][for more recent work]{tc08,cmbg08}.  Deep imaging of 38 fields of QSOs that demonstrate Mg II absorption in their spectra, combined with high resolution spectra of galaxies found in these fields, reveal the host galaxies to be mostly spiral galaxies, many with morphological asymmetries suggesting a history of mild gravitational interactions \citep{cks05,kac07,kac08}.  Some fully saturated absorption systems have been shown to correspond to Damped Lyman Alpha absorption systems \citep{rtn06} and hence to galaxies with a wide range of morphologies \citep{btj01,rao03}, whereas little is known about the hosts of the very weakest systems.  \citet{sdp94} made the first measurement of the luminosity function of Mg II host galaxies; their estimated K-band luminosity function was found to be consistent with that of \citet{mse93}, with best-fit Schechter function parameters $\phi^*=3.0 \pm 0.7 \times 10^{-2} \ (h/Mpc)^3, M^*_K = -25.1 \pm 0.3$, $\alpha=-1.0 \pm 0.3$ for a sample of 58 galaxies.  They also determined that the average absorber appears to be consistent with an Sb type galaxy (0.7 L$^*_B$), but noted a large spread  (factor of $\sim 70$) in luminosity for the sample. 

Large surveys such as the Sloan Digital Sky Survey (SDSS) have greatly increased the number of reliably detected Mg II absorption systems.  Searches for Mg II absorption systems within the spectra of SDSS QSOs have yielded close to ten thousand systems for further study \citep{ppb06,ntr05,blp06,mgiidust08}.  However, progress in more detailed analyses of these systems is hindered by the shallowness of the photometry of these QSO fields and a lack of follow-up spectroscopy, as the SDSS is limited in its ability to take spectra of objects located within 50 arcseconds on the sky of each other.  It is not feasible given current resources to carry out detailed follow up observations of thousands of fields; hence other methods must be used to gain information concerning the properties of the host galaxies of these systems and their environments. 

Recently, \citet{bmp04} and \citet{blp06} have described the use of cross-correlation techniques for studying the environments of MgII absorbers; they estimate the host halos of absorbers to be $\sim 5 \times 10^{11} M_\odot$, and find an anti-correlation between a system's measured equivalent width and the mass of the halo of its host galaxy. \citet{zib05,zib07} have considered image stacking as a way to investigate the photometric properties of MgII system host galaxies.  Their stacking analysis provides an estimate of the average luminosities and colors of these galaxies.  They conclude that the weaker absorbers are hosted by red, passively evolving galaxies, whereas the stronger absorbers are hosted by more actively star-forming galaxies.  Our study is, in some ways, complementary to theirs. Both studies use SDSS photometric data to investigate the environments of MgII systems These galaxies can then be used to determine the luminous environment of Mg II systems. Both of these methods provide a way to constrain properties of MgII system host galaxies without any follow-up observations, though our method has the advantage of being somewhat easier to implement.

In this work, we describe the results of an investigation into the absolute magnitude distribution of galaxies found near MgII absorbers.  Although we use the SDSS photometric reductions to identify galaxies around QSOs with absorption systems, these objects are generally too faint to have been part of the SDSS spectroscopic survey.  Hence, although we have colors, we do not have redshifts for these galaxies.  To compensate for the lack of redshift information, we use a background subtraction technique to isolate the galaxies physically associated with the Mg II host galaxies. We provide a discussion our sample in \S ~\ref{sample} and describe our measurement technique in \S ~\ref{method}.  We present our results in \S ~\ref{results}, and summarize our findings and their implications in \S ~\ref{discuss}. Throughout this paper we assume a $\Lambda$CDM cosmology with $\Omega_M=0.3, \Omega_{\lambda}=0.7$ and that $H_0 = 100h$~km~s$^{-1}$~Mpc$^{-1}$.


\section{The sample}\label{sample}
\subsection{The absorbers}
Our sample of Mg II absorption line systems comes from \citet{ppb06}.  Full details of the sample selection method can be found there; we only give a brief summary here.  Objects spectroscopically identified as QSOs in the SDSS Data Release 3 (DR3) are searched for evidence of Mg II absorption.  The search is confined to QSOs with $z > 0.35$.  A continuum fit for each spectrum is made using a $b$-spline to fit the underlying QSO spectrum and a principal component analysis to fit any QSO emission lines.  Spectral features are identified using a Gaussian filter method; $3.5 \sigma$ features are considered significant.  Mg II lines are identified from the resulting list of lines by looking for lines matching the doublet separation.  Features with measured equivalent width $W_{\lambda2796} > 0.8 \AA$ are compiled into the Mg II absorber sample.  From searching a total of 46420 QSOs in the SDSS DR3, there are a total of 9542 absorption systems in the final catalog.

\begin{figure}
 \centering
 \includegraphics[width=\hsize]{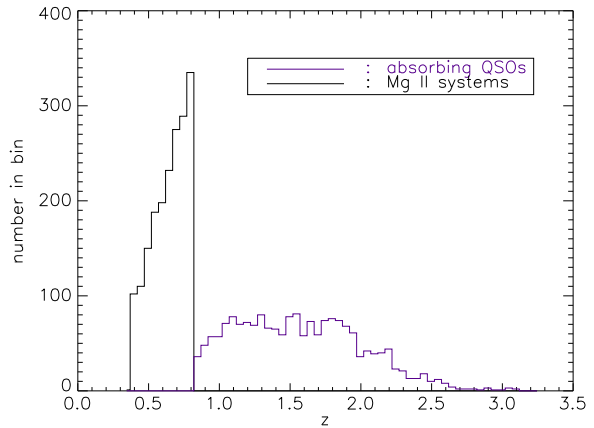}
 \caption{Redshift distributions for our final sample of Mg~II 
          absorption systems and of the background QSOs.}
 \label{dndz}
\end{figure}

The absorbers in the \citet{ppb06} sample span the equivalent width range 0.8\ \AA\ to 5.0\ \AA, with a few detections out to 10 \AA.  As Mg II absorption systems have been detected to equivalent widths of 0.02 \AA \ \citep{crcv99}, we here are investigating the properties of the strongest absorption systems. 

The redshift range over which these systems are detected is likewise broad; they are found over the full sensitivity range of the SDSS spectrograph to the Mg II doublet lines, namely $z=0.35$ to $z=2.2$.  However, the photometric catalog of the SDSS---on which we rely to study the galaxy neighbours of the Mg II host galaxies---is sensitive to galaxies out to a redshift $z \sim 1$. For this reason, we choose to concentrate on the lowest redshift absorbers.  We divide the full sample roughly into thirds and chose the lowest redshift bin for this study; we consider here absorbers in the redshift range z=0.368--0.820. A total of 2282 absorption systems fall within this redshift range.

To ensure that we accurately investigate the environment of the absorbing systems, we eliminate from our sample all QSOs that show evidence for multiple systems in their spectra.  This is necessary because we do not have redshift information for the majority of the galaxies in the neighborhood of the QSO position, and for lines of sight with multiple absorption systems it would be impossible to tell what galaxies were in the neighborhood of which absorber. This eliminates 142 systems from our sample, leaving 2140 systems.  We further eliminate all QSOs whose redshifts do not allow for possible detection of Mg II systems over the full redshift range z=0.368--0.82; this removes the lowest redshift QSOs from our sample. This is done to eliminate possible incompleteness effects in our absorption system sample. The redshift distributions for our absorption systems and the QSOs whose spectra they were found in is illustrated in figure~\ref{dndz}. Our final sample is comprised of a total of 1880 absorption systems. 

\subsection{Reference sample}
The background subtraction technique (to be detailed in \S ~\ref{method}) we will use in this paper requires the construction of a sample of random lines of sight to compare with the absorption systems.  We construct this reference sample as follows. For each QSO whose spectrum demonstrates MgII absorption (hereafter referred to as ``absorbing QSOs''), three confirmed QSOs from the SDSS DR3 which do \emph{not} demonstrate evidence for MgII absorption along their line line of sight are chosen, each having similar redshifts and $r-$magnitudes to the absorbing QSO. These QSOs shall be hereafter referred to as ``reference QSOs''. These are chosen from a set of 21543 QSOs which have $z_{\rm QSO}>0.82$ and do not have a $z>0.36$ MgII system along the line of sight.  

\begin{figure}
 \centering
 \includegraphics[width=\hsize]{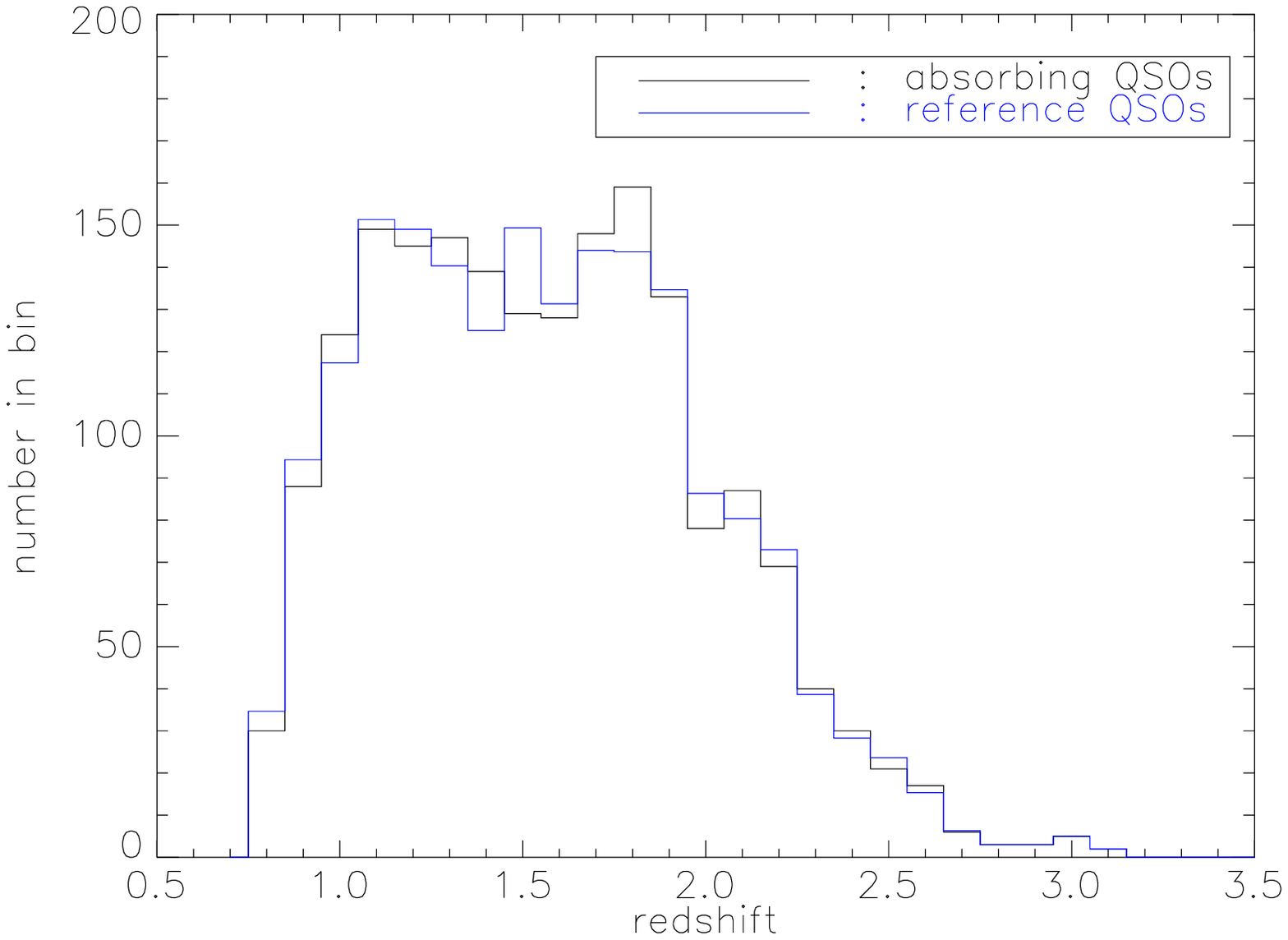}
 \includegraphics[width=\hsize]{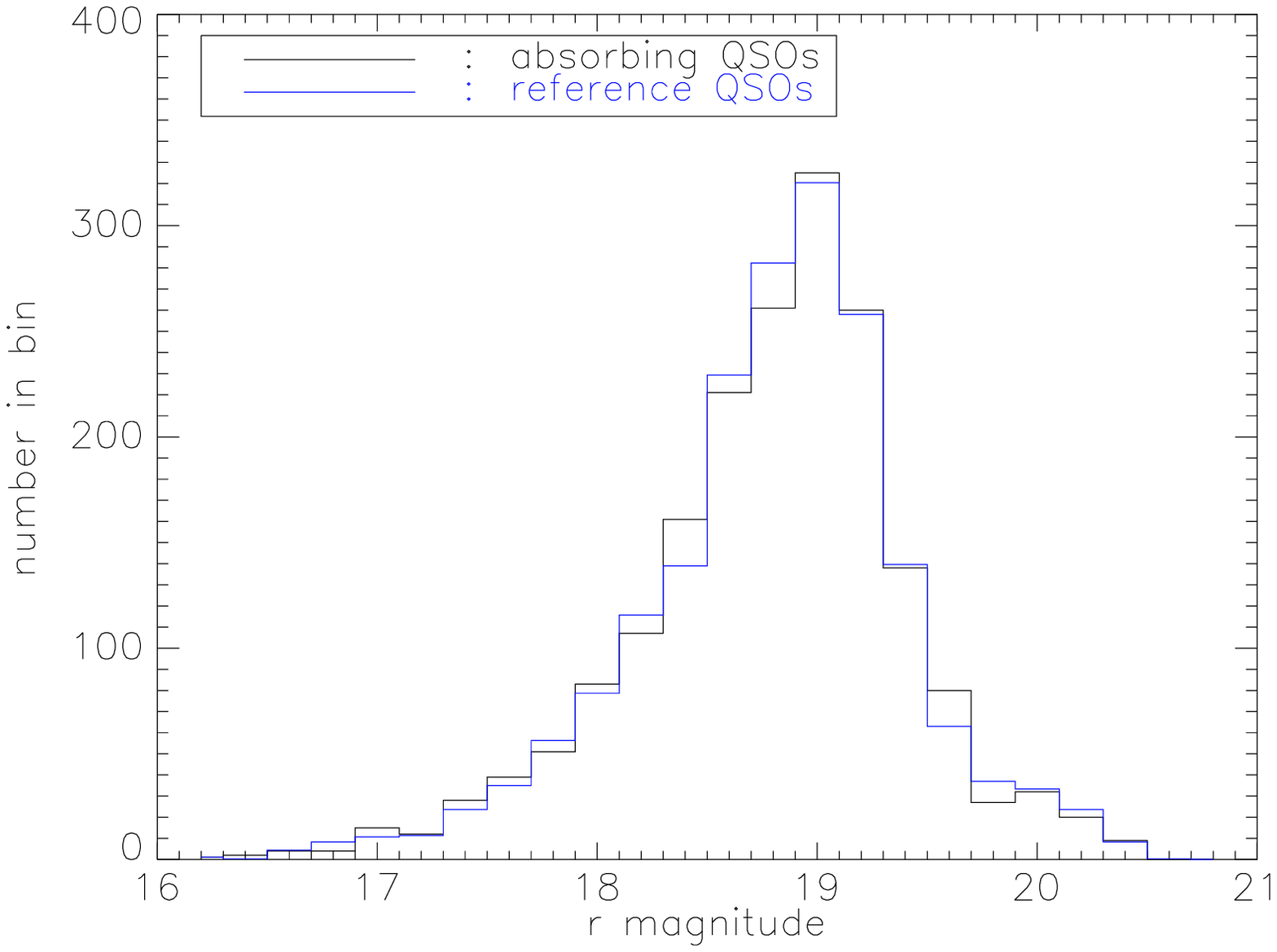}
 \caption{Redshift (top) and apparent $m_r$ magnitude (bottom) distributions 
          of the absorbing and reference QSO populations.} 
 \label{checkref}
\end{figure}

The redshift distributions of the absorbing and reference QSO populations are shown in the top panel of figure~\ref{checkref}; the $r-$magnitude distributions of the two samples are shown in the bottom panel.  The matching of redshift and r magnitude between the absorbing QSOs and reference QSOs ensures similar S/N in their spectra, as well as similar spectral coverage; that is to say, the reference and absorbing QSOs have the same redshift window over which to detect MgII, but the reference QSOs did not encounter an absorber. Each reference QSO is assigned a mock absorber whose properties are equal to those of the MgII system found along the line of sight to the absorbing QSO for which it was selected to match. As this assigned system is a ghost, its properties will be uncorrelated with galaxies found in the imaged field of the reference QSO.

The set of QSOs from which a reference sample can be drawn is about ten times larger than the sample of QSOs which contain absorbers -- the ratio of lines of sight with absorbers to those without is $1880/(1880+21543) = 0.08$.  If we include lines of sight with more than one absorption system, then this ratio is 0.085; we will return to this later.  

\section{Background subtraction technique}\label{method}

Our method for estimating the absolute magnitude distribution of galaxies neighboring Mg II absorption systems closely follows the background subtraction technique which \citet{han05} used when estimating galaxy cluster luminosity functions. We outline our technique here; a more detailed description of it can be found in the Appendix.

We begin by using the SDSS (DR3) to find those objects which are classified as galaxies and which lie within 3 arcminutes of the 1880 QSOs of our Mg II absorption sample.  We consider only galaxies with angular separations from the QSO position greater than 2 arcseconds, to eliminate any blending or seeing effects. Each galaxy is assigned the redshift of the absorption system associated with the QSO on which the field is centered. Angular separation is converted to comoving distance using a $\Lambda$CDM cosmology with $\Omega_M=0.3, \Omega_{\lambda}=0.7$. Due to the broad redshift range of our sample ($\Delta z = 0.45$), 3 arcminutes corresponds to different comoving distances from the absorber host galaxy. (For the mean redshift of our sample, $z = 0.594$, 2 arcseconds corresponds to 14.8 $h^{-1}$ kpc and 3 arcminutes to 1.33 $h^{-1}$ Mpc.) Higher redshift absorption systems sample galaxies to larger comoving separations than do lower redshift systems; thus, our sample is incomplete at these large distances.  We therefore consider only the subset of objects which lie within the range accessible over the entire redshift range: this fully sampled annulus spans comoving distances $19.3~h^{-1}$kpc $\leq d_{\rm sep}\leq 878~h^{-1}$kpc from the central QSO.

We also use the redshift of the absorber to assign absolute magnitudes to each of the galaxies in its field.  We set 
\begin{equation}
 M = m - 5 \log_{10}\left(\frac{d_{L}(z)}{10~{\rm pc}}\right) - A,
\label{distancemodeqn}
\end{equation}
where $d_L(z)$ is the luminosity distance to the galaxy and $A$ is the correction for extinction due to dust in the Milky Way from \citet{sfd98}. (Note that we do not include a k-correction term in our absolute magnitude calculations.) For those galaxies truly in the neighborhood of the Mg II absorption system, this procedure yields the true absolute magnitude.  It of course yields an incorrect magnitude for all the other objects.  

We then follow the same procedure for each of the 5640 reference QSO positions:  Galaxies projected within 3 arcminutes of each reference QSO are found and assigned redshifts as described above.  Their angular separation from the reference QSO's position are converted to comoving distances based on the redshift of the ghost absorber, and galaxies located within the fully sampled annulus are kept.  These are assigned absolute magnitudes on the basis of their (ghost absorber) redshift.  In this case, essentially all distances and luminosities are incorrect.  

We now have absolute magnitude distributions centered on the absorber and reference populations.  For the absorber population,
\begin{equation}
 N_{\rm absorber}(M) = N_{\rm neighbors}(M) + N_{\rm random}(M);
\end{equation}
whereas for the reference population,
\begin{equation}
 N_{\rm reference}(M) = N_{\rm random}(M).
\end{equation}
Here $N_{\rm absorber}(M)$ denotes the number of galaxies with absolute magnitude M found in the field of an absorbing QSO, $N_{\rm random}(M)$ the number of such galaxies randomly projected into the field, and $N_{\rm neighbors}(M)$ the true neighbours of the absorption system host.  If we subtract these two distributions---the absorber and reference distributions--taking care to account for the fact that we have three times as many QSOs in the reference catalog as in the absorber one, all that will remain is the contribution to the counts from those galaxies which are the true neighbors of the absorption system.  {\em These are precisely those objects for which distances and absolute magnitudes were appropriately calculated}.  Hence, the absolute magnitude distribution of MgII system neighbour galaxies can be determined from this calculated difference in measured distributions.  

The Appendix provides a detailed calculation showing that the quantity which the background subtraction technique actually estimates is
\begin{equation}
 N_{\rm neighbours}(M) \approx V_{\xi} \int {\rm d}z_{\rm abs}\, 
                                  \frac{{\rm d}N}{{\rm d}z_{\rm abs}}
                                  \,\phi(M|z_{\rm abs});
 \label{NnbrsM}
\end{equation}
here $\phi(M|z_{\rm abs})$ is the luminosity function in fields of 
effective volume $V_{\xi}$ centered on absorbers 
(c.f. equation~\ref{xicorrnofbigm}).  The Appendix also describes the results of testing our procedure on a mock catalog of galaxies, subjected to the same ``observing" limits as the SDSS, to ensure that it does in fact recover the absolute magnitude distribution of MgII absorber neighbours.  

Our background subtraction technique relies on the existence of an over density of galaxies around absorption systems which is not present in a random sample. Because galaxies cluster over scales of $\approx$ 100 kpc---10,000 kpc, such an over density of galaxies is expected. Hence, galaxy clustering ensures the viability of our technique. However, we have not yet addressed the question of how many lines of sight are required to make a statistically significant measurement.  If the number of galaxies correlated with absorbers is some fraction $C$ of the background counts, $N_{\rm absorber}(M) = (1+C)\, N_{\rm random}(M)$.  Then the `signal' in our background subtraction method is $C\, N_{\rm random}(M)$, where $N_{\rm random}$ is proportional to the number of lines of sight.  If the number of reference QSOs is many times that of absorbing QSOs, or if the reference population was calculated analytically (c.f. equation \ref{nofbigmmc}), then the Poisson noise on our measurement would be $\sqrt{(1+C) N_{\rm random}(M)}$.  So if we want $C\, N_{\rm random}(M)\ge 3\,\sqrt{(1+C) N_{\rm random}(M)}$ then the required sample size is  
\begin{equation}
 N_{\rm random}(M) \ge 3^2\frac{(1+C)}{C^2}.
 \label{SN}
\end{equation}
This shows that our technique requires $N_{\rm random}(M)\gg 1$ if $C\ll 1$.  For example, if $C \approx 10\%$, then $N_{\rm random}\approx 10^3$.  If there are only a few galaxies per line of sight, then our technique requires many lines of sight if $C\ll 1$.  If there are many galaxies per line of sight (e.g. if MgII absorbers were centered on massive galaxy clusters), then fewer lines of sight would be required.  

Above, we noted that we did not include a k-correction term when calculating absolute magnitudes for galaxies in our sample. As a result, our observed $r-$magnitudes do not correspond to the rest-frame $r-$magnitudes of MgII system neighbour galaxies. For most of the absorption systems in our sample, the $r-$band absolute magnitudes we calculate are close to rest-frame B-band absolute magnitudes; systems with $z > 0.7$ are closer to rest-frame U. The SDSS is not expected to find many galaxies with $z>0.7$, so we do not expect to find many MgII system neighbour galaxies in this redshift range. Thus including them in our survey should not contaminate our results very much.

Given the background subtracted absolute magnitude distribution, we could calculate a luminosity function once we account for the fact that our galaxy survey is magnitude limited, so more luminous galaxies are seen to greater distances.  This could be done by carefully implementing Schmidt's $V_{\rm max}$ method for the particulars of our survey (see, e.g., Appendix).  In practice, because our sample spans a reasonable range in $z$, and we do not apply k-corrections, we do not really distinguish k-corrections from evolution; therefore, estimating the luminosity function using such a $V_{\rm max}$ method is slightly tricky.  For this reason, we present our results in terms of the observed absolute magnitude distribution, $N_{\rm neighbours}(M)$, rather than an evolution corrected luminosity function.  

\begin{figure}
 \centering
 \includegraphics[width=0.95\hsize]{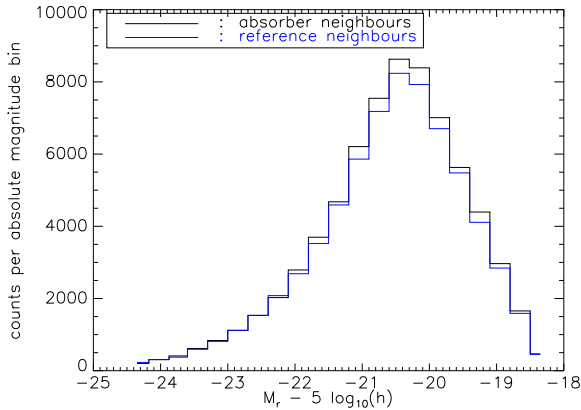}
 \caption{Absolute magnitude distributions for the absorber population, plotted in black, and the reference population, plotted in blue.
 }
 \label{totaldNdM}
\end{figure}

\section{Results}\label{results}

Before carrying out the background subtraction procedure detailed in \S ~\ref{method}, we first present the absolute magnitude distributions of the absorber and reference populations in figure~\ref{totaldNdM} to determine if there is a non-negligible difference between the two. Such a difference would correspond to the true neighbours of the MgII system host; this is the population we aim to isolate. Figure~\ref{totaldNdM} shows that there is an excess of absorber population counts over reference population counts from $-22.0 \leq M_r \leq -18.0$. This indicates, even before our method has been applied, that we have successfully found an over density of galaxies in the absorber population, not extant in the reference population, which are  associated with the absorption systems in their fields.  These galaxies are all that should remain upon implementation of our background subtraction technique. 

\subsection{Reference sample}
\begin{figure}
 \centering
 \includegraphics[width=\hsize]{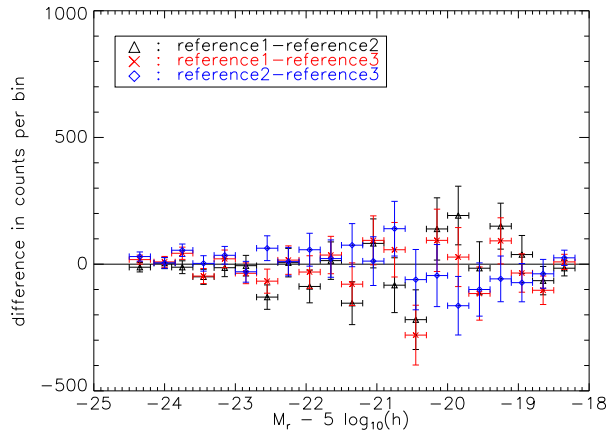}
 \caption{Difference between the reference samples. 
          A line at zero counts is shown for reference.}
 \label{diffref}
\end{figure}

We test that our reference population is truly comprised of galaxies randomly projected into the fields of the central reference QSOs before implementing our technique. To do so, we split our reference catalog into three equal-sized catalogs, each containing one reference QSO for every absorbing QSO. We then subtract the mutual absolute magnitude counts for each catalog pair to see if the resulting distribution is centered on zero to within the scatter. Figure~\ref{diffref} shows the various pairwise differences between the three reference catalogs. The absence of any real feature in this figure is reassuring; it suggests that our reference sample is truly random.  

\begin{figure}
 \centering
 \includegraphics[width=\hsize]{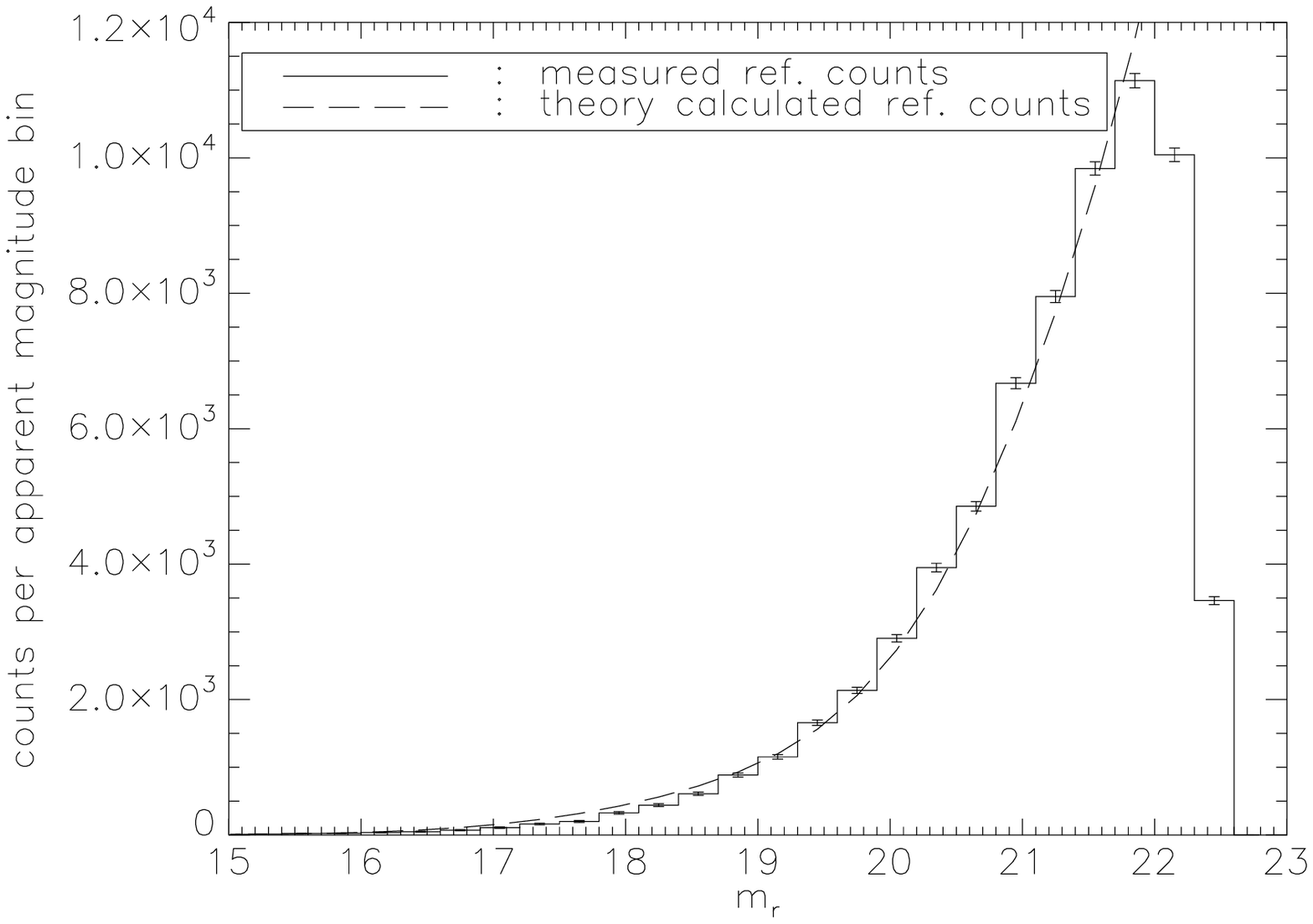}
 \includegraphics[width=\hsize]{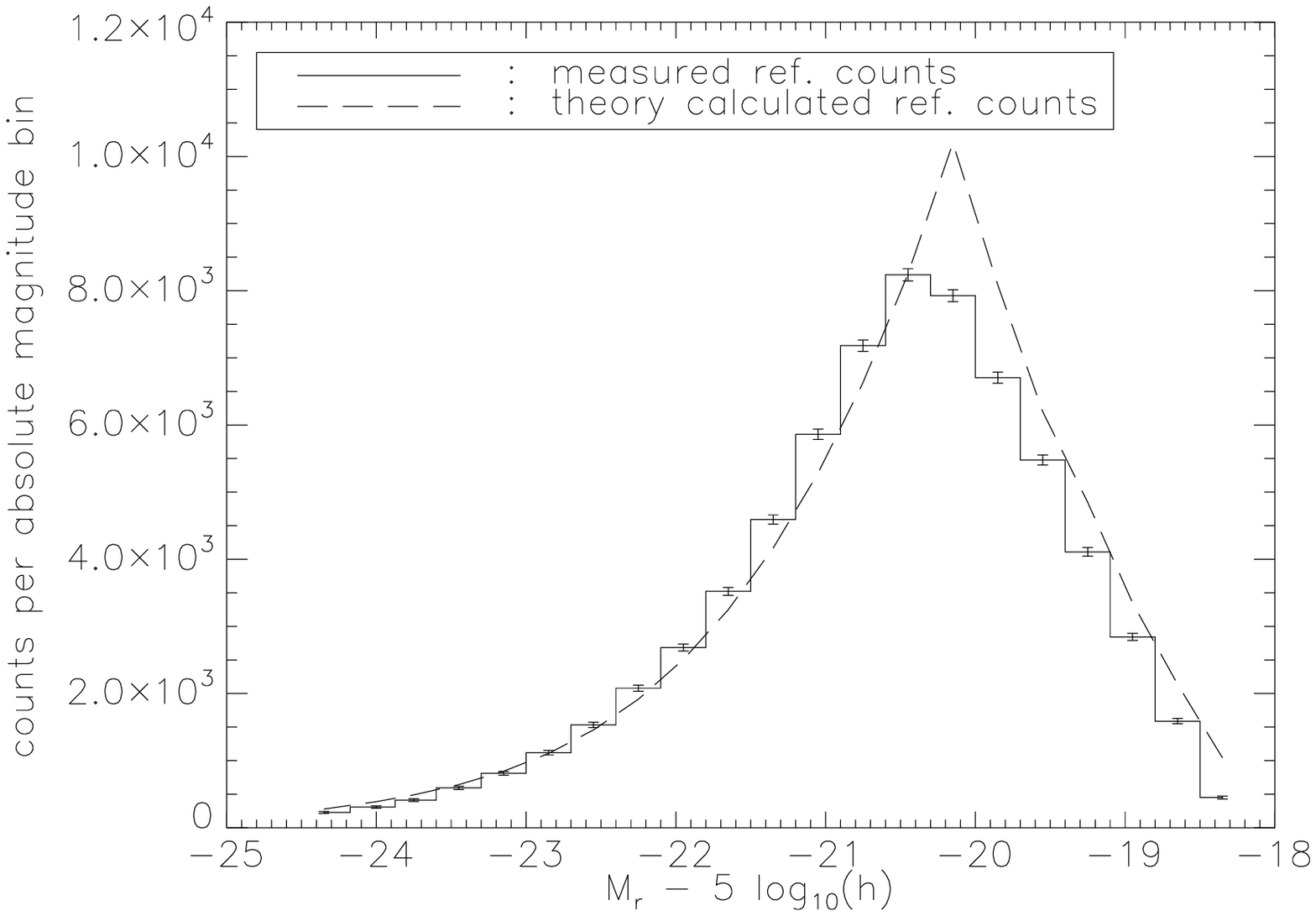}
 \caption{Apparent (top) and absolute (bottom) magnitude distributions (histograms). The dashed lines show predictions based on equations~\eqref{noflittlem} and~\eqref{nofbigmsub} for the top and bottom panels respectively.}
 \label{theoryref}
\end{figure}

We perform one additional test on our reference sample to ensure its random nature. As described in the Appendix, the expected distribution of apparent magnitudes in a randomly chosen field can be computed analytically if the luminosity function is known (equation~\ref{noflittlem}).  The result of assuming that this is given by the COMBO-17 survey \citep{wolf03} and the ESP survey \citep{zuc97} (converted to our cosmology following \citet{lisk03}) is shown by the dashed curve in the top panel of Figure~\ref{theoryref}.  (These surveys span the same redshift range as our dataset.)  The histogram shows the actual counts; while our calculated apparent magnitude distribution matches the observed one through $m_{r} \approx 21.5$, we over-estimate the number of galaxies with $m_{r} > 21.5$.  This is due, in part, to our neglect of k-corrections, which have the effect of smearing out what would otherwise be a sharp cut at $m_r$.  The reference sample for each absorber is constructed by shifting this apparent magnitude distribution by factor which depends on $z_{\rm abs}$.  Since the distribution of $z_{\rm abs}$ is known, we can convert this apparent magnitude distribution to $N_{\rm reference}(M)$ (see equation~\ref{nofbigmsub}).  The dashed line in the bottom panel of Figure~\ref{theoryref} shows the result of transforming the dashed curve in the upper panel in this way (i.e., using equation~\ref{nofbigmsub}), and the histogram shows the actual distribution of $N_{\rm reference}(M)$.  The small differences between the predicted and actual distributions can be traced to the differences in the top panel.  Given our understanding of the nature of these differences, and the overall agreement in shape between the calculated and observed distributions, we conclude that our reference population is indeed consistent with a population of galaxies randomly projected into a field.

\subsection{Background subtraction and the full sample}
\begin{figure}
 \centering
 \includegraphics[width=\hsize]{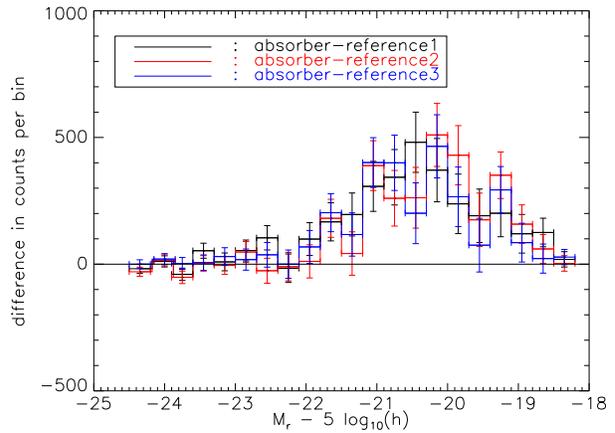}
 \caption{Difference in observed counts between the absorber sample and 
          each of the three reference samples.}
 \label{diffnm}
\end{figure}

We now implement the procedure outlined in \S~\ref{method} to isolate MgII system neighbour galaxies.  Figure~\ref{diffnm} shows $N_{\rm absorber}(M)-N_{\rm reference}(M)$ as a function of $M$, for each of the three equal-sized reference catalogs we used to test our reference sample (see Figure~\ref{diffref}).  In all cases, there is a broad well-defined bump centered on $M_r \approx -20.2$.  That we see a bump of approximately the same shape peaking at approximately the same place in all three distributions indicates that we measure a real underlying one and not a by-product of noise.  Moreover, the bell-like shape is similar to that which one typically obtains if the raw galaxy counts in an apparent magnitude limited survey are plotted as a function of absolute magnitude.  This gives us confidence that our method has worked.  In what follows, we will reduce the noise in our estimate of the background counts by using the full reference catalog compiled in \S ~\ref{method} when subtracting, rather than the three equal-sized catalogs.  When we do so, we detect a total of 2797 galaxies --- about 1.5 times the number of absorbers.  

Because of luminosity evolution in our galaxy population over the range $0.37 \leq z \leq 0.82$, further complicated by potential confusion with k-corrections (which we ignore), we do not convert our background subtracted counts into an estimate the luminosity function and show the resulting $\phi(M)$.  Rather, we compare these counts directly with predictions based on inserting various models for the luminosity function and its evolution into equation~\eqref{NnbrsM}.  When doing so, we treat $V_\xi$ as a free parameter; recall that this parameter is a measure of the effective volume associated with absorbers.  

As we did when modelling the reference sample, we use luminosity functions from the COMBO-17 survey to model our background subtracted sample.  Since we do not k-correct our absolute magnitudes, our observed $r-$magnitudes correspond to their rest-frame B-magnitudes up to $z \approx 0.7$, covering almost the entire redshift range of galaxies considered (as noted above). The redshift range of this survey is well-matched to that of the galaxies we detect with our method; galaxy luminosity functions for four different populations are reported therein for mean redshifts \={z}=0.3, \={z}=0.5, \={z}=0.7, \={z}=0.9, and \={z}=1.1. This amply covers the redshift range over which we detect MgII neighbor galaxies.

\begin{figure}
 \centering
 \includegraphics[width=0.95\hsize]{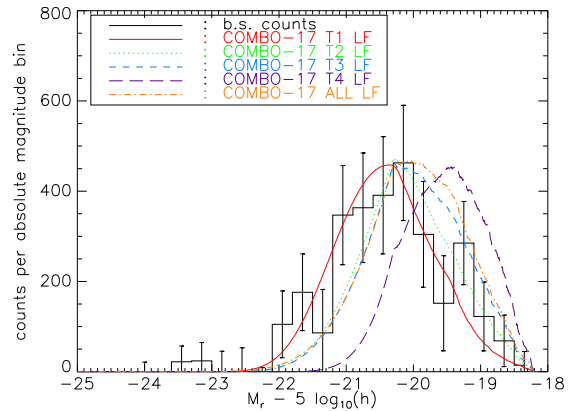}
 \caption{Background subtracted absolute magnitude distribution of MgII system neighbour galaxies (histogram) and expectations based on inserting luminosity functions from the COMBO-17 survey into equation~\eqref{NnbrsM}.}
 \label{dNdMlitcompare}
\end{figure}

As noted above, COMBO-17 reports luminosity functions for four galaxy types, as well as an estimate for the full population.  In our estimate of the reference counts, we used only the full population.  Here, to explore the possibility that MgII absorption system neighbors may be preferentially of one type, we show the result of inserting each of the four luminosity functions, as well as that for the full population, into equation~\eqref{NnbrsM}.  The four smooth curves in Figure~\ref{dNdMlitcompare} show these different predictions; in each case the overall amplitude $V_\xi$ has been left as a free parameter.  The histogram shows the background subtracted counts that we measure.  Evidently, the COMBO-17 Type 1 luminosity function provides the best description of our measurements.  Both the Type 2 and Type 3 luminosity functions provide an acceptable fit to the faint-$M$ end of the distribution, but neither match well at the bright end. The same is true for the total (i.e. all types) luminosity function. The Type 4 luminosity function provides the poorest fit to the data.

We conclude that our observed distribution seems to be most consistent with that calculated for a COMBO-17 Type 1 luminosity function, which according to \citet{wolf03} is drawn from galaxy types ranging from E---Sa. Taken at face value, then, our comparison suggests that the neighbours of MgII absorption systems are unlikely to be very late-type galaxies.  It is interesting to note, in passing, that we find that we require $V_\xi \approx 355$ to fit our background subtracted counts when the Type 1 luminosity function is considered.  This in equation~\eqref{NxiFabs} suggests a correlation length $r_0$ of about 4.6~$h^{-1}$Mpc, which is not unreasonable.  

\subsection{Samples split by equivalent width}
To investigate the possibility that absorption systems of different strengths are associated with different environments, we divide our sample in half according to equivalent width.  The dividing point occurs at rest-frame equivalent width (REW) 1.28\AA.  Hereafter, we shall refer to the sub-sample with 0.88 \AA $\leq$ REW $\leq$ 1.28 \AA \ as the ``weak'' sample, and the sub-sample with REW $>$ 1.28 \AA \ as the ``strong'' sample.  Note that, when we divide the sample, each absorber QSO ``keeps" its three reference QSOs; this ensures that the redshift and apparent magnitude distributions of our reference samples match those of the subsamples.   

\begin{figure}
 \centering
 \includegraphics[width=0.95\hsize]{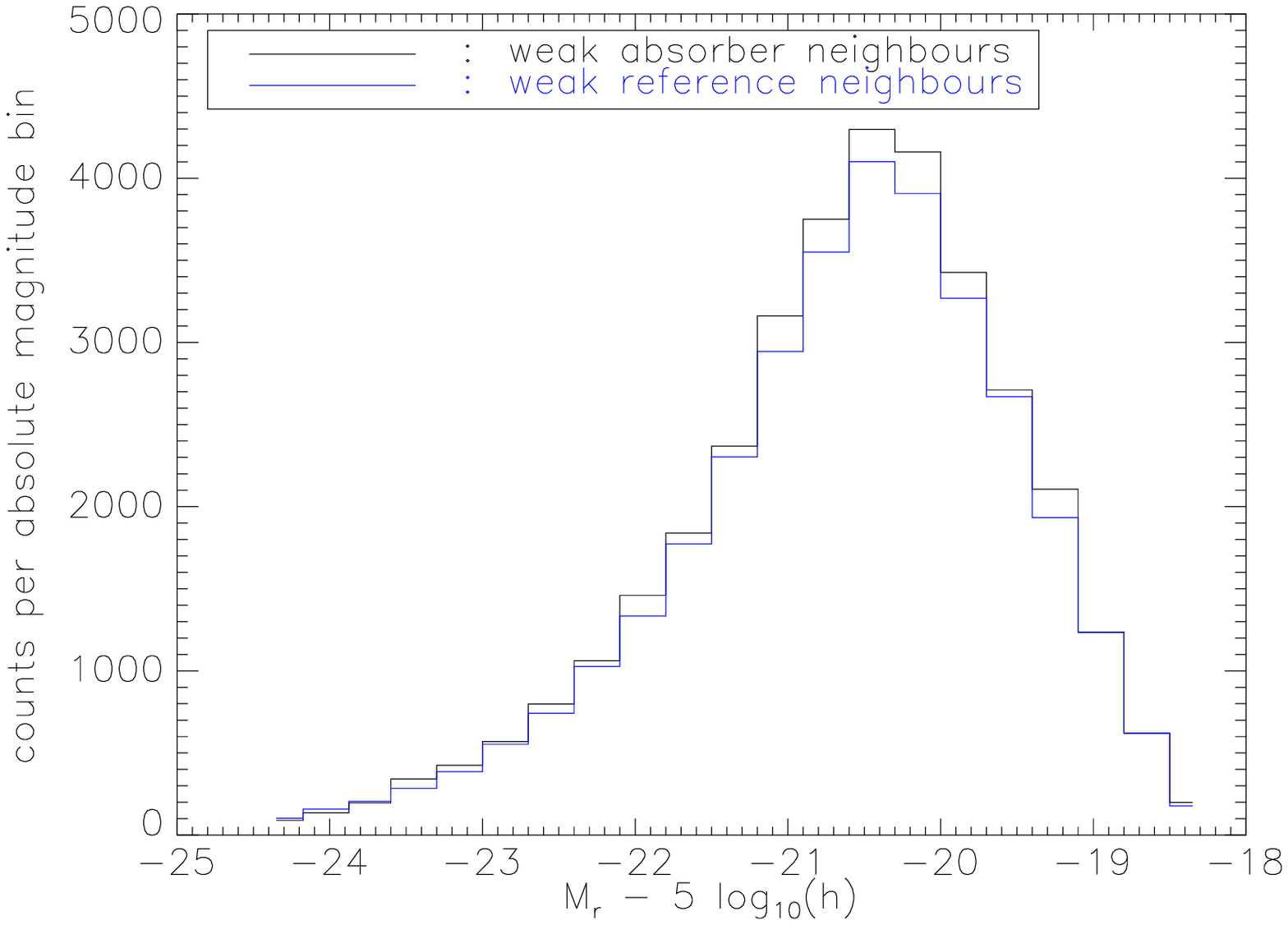} 
 \includegraphics[width=0.95\hsize]{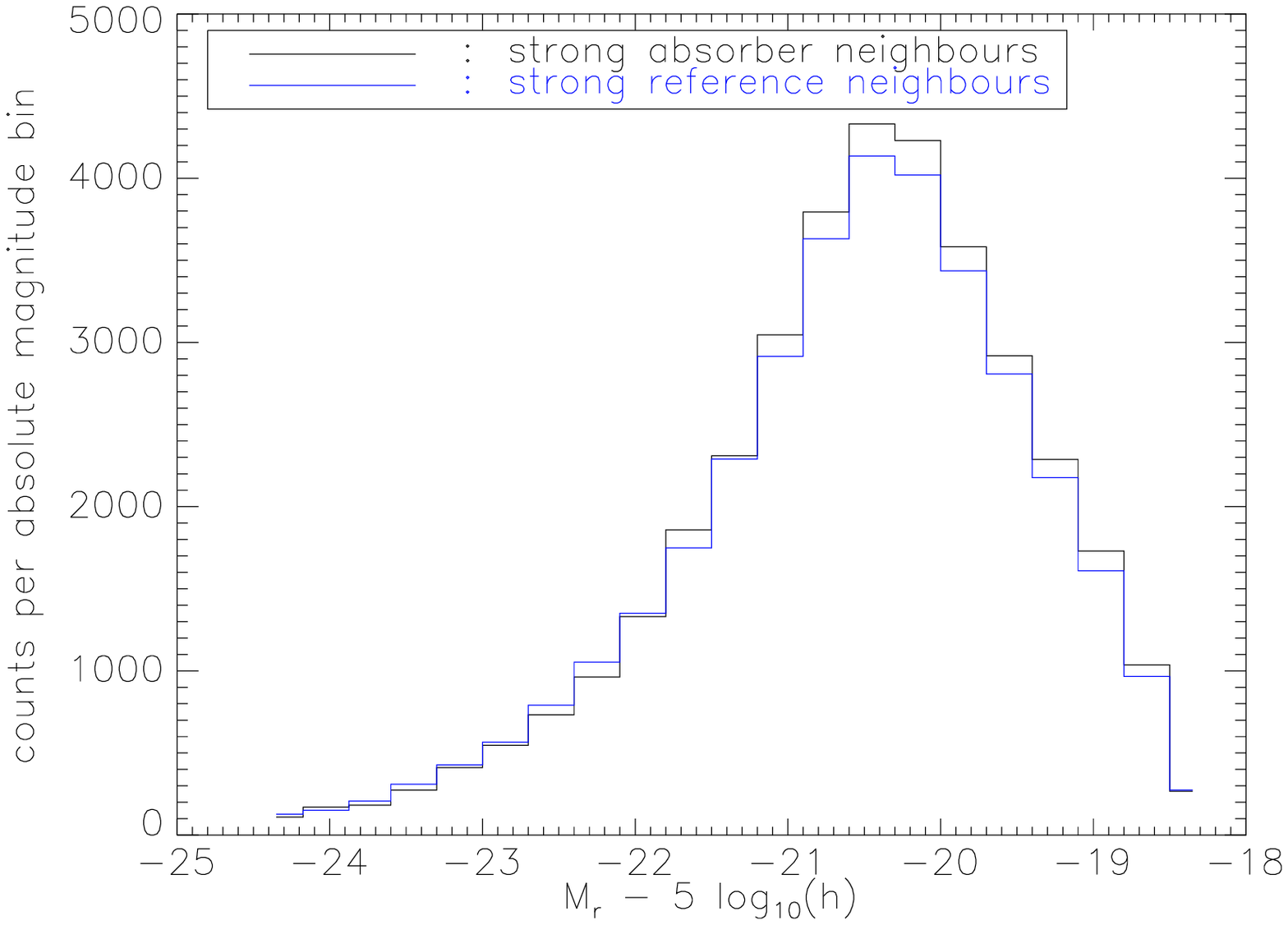}
 \caption{Absolute magnitude distributions for the weak (top) and strong (bottom) absorber and reference populations.}
 \label{splitdNdM}
\end{figure}

As we did with the full population (compare Figure~\ref{totaldNdM}), we first present the absolute magnitude distributions of the weak and strong absorber and reference populations.  Both the top and bottom panels of Figure~\ref{splitdNdM} show an excess of absorber population counts over reference population counts over the range $-22.0 \leq M_r \leq -19.0$.  This indicates that we find neighbours around both weak and strong absorption systems rather than around just one sub-sample.  

\begin{figure}
 \centering
 \includegraphics[width=0.95\hsize]{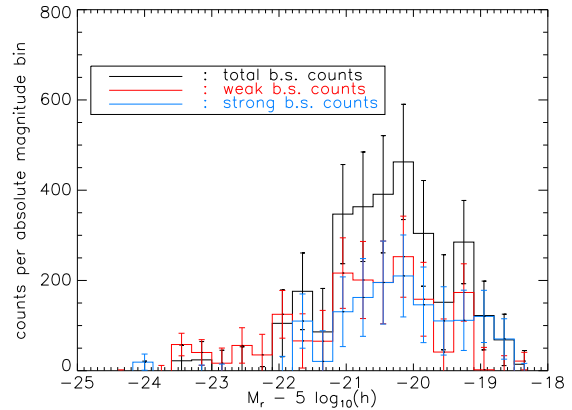}
 \caption{Background subtracted absolute magnitude distributions of the weak and strong sub-samples. The background subtracted absolute magnitude distribution of the full sample is also shown for reference.}
 \label{weakandstrongdNdM}
\end{figure}

The light-grey and dark-grey histograms in Figure~\ref{weakandstrongdNdM} show the result of subtracting reference from absorber counts in the weak and strong samples, respectively.  Note that, because of how we have defined our reference samples (i.e., each absorber QSO ``keeps" its three reference QSOs), the background subtracted counts sum to give the absolute magnitude distribution of the total sample (black histogram, same as shown in Figure~\ref{dNdMlitcompare}), by definition.  Both distributions are clearly different from zero; they both show a broad bump over the absolute magnitude range $-19.0 \leq M_{r} \leq -22.0$. Both the weak and strong distributions appear to peak at $M_{r} \approx -20.2$, though the peak of the strong distribution is slightly better defined than the peak of the weak distribution. 

The two distributions do, however, appear to be slightly offset; the strong distribution extends from $-18.0 \leq M_{r} \leq -22.0$, whereas the weak distribution extends from $-19.0 \leq M_{r} \leq -23.5$. The weak distribution also features a high-M tail which is not present in the strong one. As another indication of the difference in distributions, note that the background subtracted absolute magnitude distribution for the weak sample contains 1679 galaxies, whereas that for the strong sample contains 1118 galaxies -- even though the two samples were centered on the same number of absorbers: 940.  This suggests that galaxies near weak systems are more readily detectable by the SDSS than are galaxies near stronger systems, which would imply that weaker systems are surrounded by brighter galaxies provided both sub-samples have similar redshift distributions.

\begin{figure}
 \centering
 \includegraphics[width=0.95\hsize]{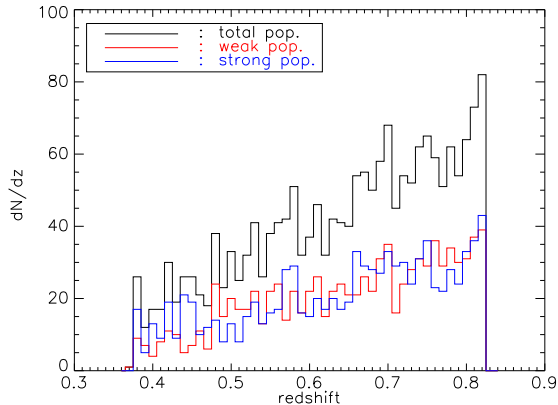}
 \caption{Redshift distribution of weak (red) and strong (blue) MgII absorption systems.}
 \label{absorberdndz}
\end{figure}

If the redshift distributions of the two absorber populations were the same, we could conclude from the absolute magnitude distributions shown in Figure~\ref{weakandstrongdNdM} that the luminosity function of galaxies centered on strong absorbers includes more faint objects.  Figure~\ref{absorberdndz} suggests that $dN/dz$ is indeed similar for the two populations.  As an extra check, we performed the same analysis as before -- using equation~\eqref{xicorrnofbigm} to calculate absolute magnitude distributions from COMBO-17 luminosity functions, using the redshift distribution appropriate for each sub-sample, and leaving $V_\xi$ as a free parameter.  

\begin{figure}
 \centering
 \includegraphics[width=0.95\hsize]{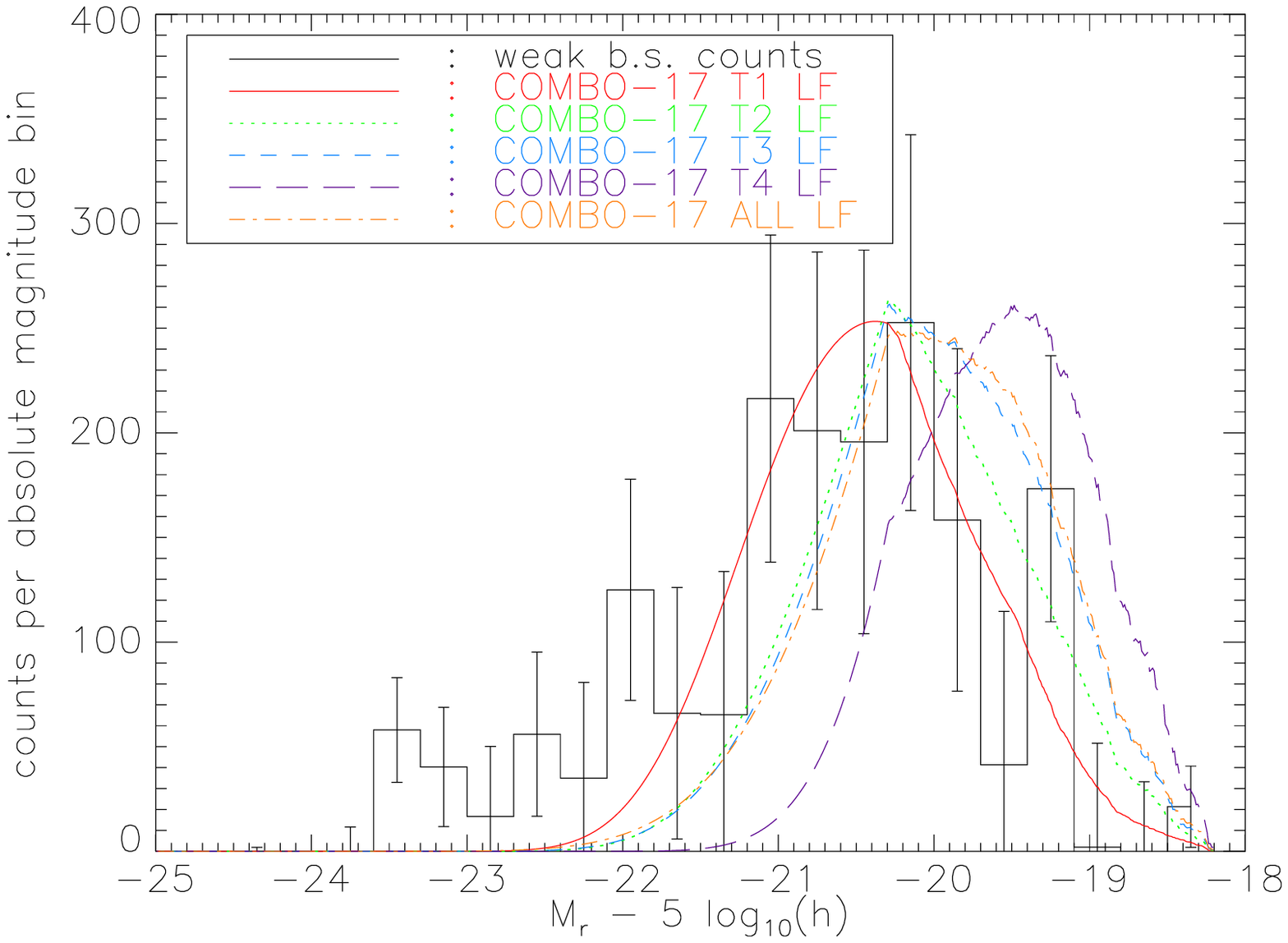}
 \includegraphics[width=0.95\hsize]{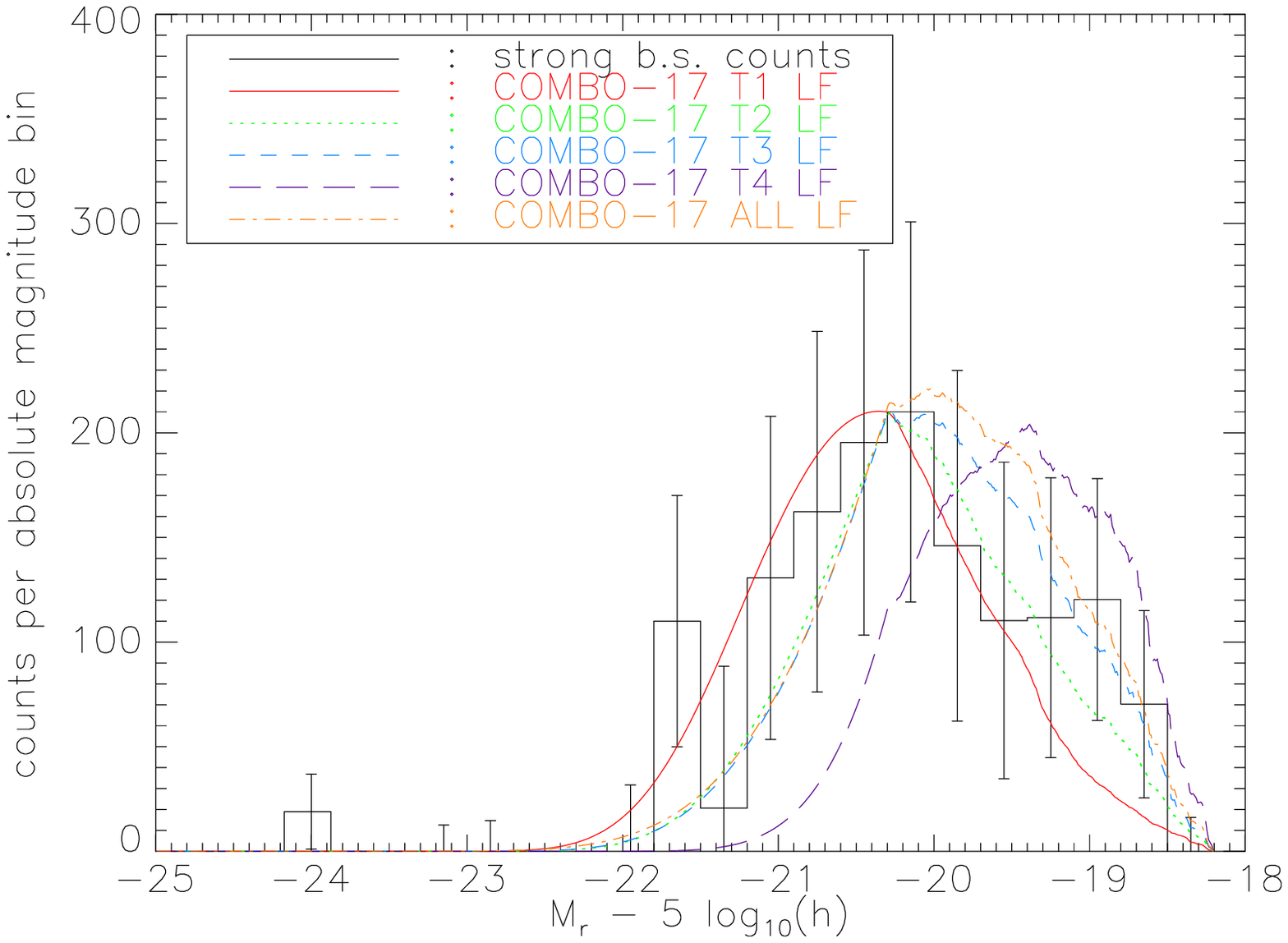}
 \caption{Background subtracted absolute magnitude distributions for the weak (top) and strong (bottom) sub-samples, compared with the expected distribution based on various luminosity functions from the COMBO-17 survey.}
 \label{weakstrongdNdMlitcompare}
\end{figure}

The top panel in Figure~\ref{weakstrongdNdMlitcompare} shows results for the weak absorber population; a COMBO-17 Type~1 luminosity function provides a good description of our measurements, at least over the range $ -22 \leq M \leq -18$.  The bottom panel shows that although a Type~1 luminosity function also provides the best description of the bright end, there is a noticeable excess at low luminosities, which could be described by a Type 4 luminosity function.  Thus, it appears that whereas weak absorbers tend to not be associated with late-type galaxies, at least some strong absorbers may well be.  This is consistent with the results of \citet{zib07} and \citet{blp06}, who used very different techniques from ours.  

\section{Discussion}\label{discuss}

\subsection{Summary of results}
We have estimated the absolute magnitude distribution of galaxies which lie within about 1~$h^{-1}$Mpc of MgII absorption line systems.  Our sample of 1880 absorbers, which is drawn from the SDSS DR3 MgII catalog of \citet{ppb06}, spans the redshift range $0.368 \leq z \leq 0.820$, and consists of systems with rest-frame equivalent width REW $>$ 0.88\AA.  Lines of sight demonstrating MgII absorption at multiple redshifts have been eliminated, as have QSOs whose spectra would not allow systems to be detected over the entire redshift range.

Most of the galaxies in SDSS imaging which lie within about 1~$h^{-1}$Mpc of each of these objects have five band imaging but no spectra.  Hence their redshifts are not known; this, of course, complicates estimates of the absolute magnitude distribution.  In principle, we could estimate luminosities using photometric redshifts; \citet{she07} describes how to derive accurate estimates of the luminosity function from photo-$z$s.  In the present context, however, we use only the photometric information for the galaxies, because we {\em do} know the redshift of the absorber.  Hence, we use this information and a background subtraction technique to statistically remove foreground and background galaxies.  

Galaxies located within about 3 arcminutes of absorbing QSO positions in the SDSS DR3 imaging are assigned the redshift of the absorption system; their absolute magnitudes are calculated based on this redshift.  To correct for foreground and background galaxies that have been projected into the field, we carry out a similar procedure on a reference set of QSOs.  The reference QSOs have the same redshift and $r$-band magnitudes as the QSOs with MgII absorption features (Figure~\ref{checkref}), but do not demonstrate absorption in their spectra.  Galaxies in the fields surrounding a given reference QSOs are assigned the same redshift as the QSO with absorption features which that reference QSO was chosen to represent. A detailed analytic description of the method is provided in Appendix~A1.  Tests of our technique on a mock catalog of galaxies that is subjected to a similar reduction process as was used for the actual data indicate that it works well (Appendix~\ref{mocks}).

The background-subtracted absolute magnitude distribution we observe peaks at $M_r \approx -20.2$ and is roughly bell-shaped over the range $-22.0 \leq M_r \leq -18.0$ (Figure~\ref{diffnm}). Comparison with expectations from published luminosity functions over this same redshift range (from COMBO-17) suggests that the galaxies within $\sim 1 \ h^{-1}$Mpc of fields centered on absorbers are of Type~1, meaning they may be ellipticals or Sa's (Figure~\ref{dNdMlitcompare}).  Subdivision into weak and strong systems suggests that late-type galaxies may be associated with the stronger systems (Figure~\ref{weakstrongdNdMlitcompare}), but are most likely not associated with the weaker systems, in agreement with previous work based on very different methods \citep{blp06, zib07}.

\subsection{Implications for covering fraction}\label{fcover}
The fraction of eligible lines of sight which contain an absorber is about 8 percent.  The galaxies which our background subtraction technique identifies are the neighbours of absorption systems. If we assume that at least some of these neighbours are in fact the MgII system hosts themselves, then this observed fraction can be used to place interesting constraints on the sizes of absorbers as follows. (Note that this is not the main focus of our paper).  If absorbers are associated with galaxies, and are sufficiently rare such that they do not overlap, then the area on the sky they cover is 
\begin{equation}
 \int dz\,\frac{dV}{dz}\, \int dL\, \phi(L|z)\,\kappa(L,z)\,
  \pi \left(\frac{R(L,z)}{d_A(z)}\right)^2,
\end{equation}
where $\phi(L|z)$ is the luminosity function, $d_A(z)$ is the angular diameter distance, $R(L,z)$ is the radius out to which MgII absorption is seen in the observed range of equivalent widths (e.g., some models have equivalent width decreasing with distance from the center of the galaxy, with a normalization which depends on $L$), and $\kappa$ is the fraction of galaxies which have an absorber (e.g., if the absorbers are small clouds within a galaxy, or if only a fraction of galaxies actually contain MgII).  

It is common to parametrize 
\begin{equation}
 R(L,z) = R_{*W} \left(\frac{L}{L_*(z)}\right)^\beta,
 \label{sizeL}
\end{equation}
and to assume $\kappa$ is a constant \citet{sdp94}. Recent work suggests that $\beta\approx 0.35$ \citep{ct08}, or $\beta\approx 0.2$ \citep{kac08}.  Then, for a Schechter luminosity function with faint-end slope $\alpha$, the expected covering fraction for our sample is:
\begin{eqnarray}
 F &=& \kappa \, (c/H_0)  \, (\pi R^2_{*W})\ \Gamma(1+\alpha+2\beta) \nonumber\\
    &&\qquad\times\quad \int_{0.37}^{0.82} dz\,\frac{\phi_*(z) \,(1+z)^2}
                             {\sqrt{\Omega_0(1+z)^3 + (1-\Omega_0)}}.
 \label{fintegral}
 \end{eqnarray}
If the galaxies isolated by our background subtraction method are drawn from a COMBO-17 Type 1 $z=0.5$ luminosity function ($\phi_* = 0.0028\, h^3$Mpc$^{-3}$ and $\alpha=0.52$) and we ignore evolution, then the expected covering fraction is
\begin{equation}
 F = 0.25 \,\kappa \left(\frac{R_{*W}}{100h^{-1}~{\rm kpc}}\right)^2.
\label{coverfractiontype1}
\end{equation}
We have used $\beta=0.35$ in our calculations. Setting  $\beta=0.2$ decreases the right hand side by ten percent.  Since we find $F=0.08$, $\kappa = 0.32$ if $R_{*W}=100h^{-1}~{\rm kpc}$, and $R_{*W}\approx 57h^{-1}~{\rm kpc}$ if $\kappa = 1$.  We can account for luminosity evolution in our calculation of equation \eqref{fintegral} by using the Type 1 luminosity functions reported by Wolf et al. (2003) for redshifts \={z}=0.3, \={z}=0.5, \={z}=0.7, and assuming no evolution between $0.3 \leq z < 0.5$, $0.5 \leq z < 0.7$, and $0.7 \leq z < 0.9$. Doing so, we find that
\begin{equation}
 F = 0.18 \,\kappa \left(\frac{R_{*W}}{100h^{-1}~{\rm kpc}}\right)^2.
\label{cfractiontype1}
\end{equation}
For $F=0.08$ and $R_{*W}=100h^{-1}~{\rm kpc}$, $\kappa = 0.44$ is required; if instead $\kappa = 1$ then $R_{*W}\approx 67h^{-1}~{\rm kpc}$. The values of $\kappa$ we obtain with our rough estimates are lower than those found in the literature, but not unreasonably so. Similarly, our estimate of $R_{*W}$ is reasonable. We note, however, that for a galaxy with REW $\geq 0.88$, a size of $\approx 70 h^{-1}~{\rm kpc}$ is larger than would be expected from Figure 4 of \citet{ct08}.

To illustrate that the estimated $\kappa$ and $R_{*W}$ depend on the adopted luminosity function, suppose that these galaxies are drawn from the COMBO-17 total luminosity function ($\phi_* = 0.018\, h^3$Mpc$^{-3}$ and $\alpha=-1.1$ at $z=0.5$).  Then
\begin{equation}
 F = 2.05 \,\kappa \left(\frac{R_{*W}}{100h^{-1}~{\rm kpc}}\right)^2
\end{equation}
if we ignore evolution.  Setting $\beta=0.2$ approximately doubles the right hand side.  With $F=0.08$, $R_{*W}=100h^{-1}~{\rm kpc}$ implies $\kappa = 0.039$, and $R_{*W}\approx 19.75h^{-1}~{\rm kpc}$ if $\kappa = 1$.  Accounting for evolution as before (in this case $\alpha$ also evolves, so we keep the Gamma function piece inside the redshift integral) we find 
\begin{equation}
 F = 3.03 \,\kappa \left(\frac{R_{*W}}{100h^{-1}~{\rm kpc}}\right)^2,
\end{equation}
implying $\kappa=0.026$ if $R_{*W}\approx 100h^{-1}~{\rm kpc}$ and $R_{*W}\approx 16.25h^{-1}~{\rm kpc}$ if $\kappa=1$. These values are illustrative only, because this luminosity function does {\em not} result in good agreement with our background subtracted counts.  

\subsection{Future work}

Our background technique is generally applicable to other studies in which redshifts are known for only a small subset of objects, but these are known to be correlated with a larger sample, for which we wish to estimate luminosities.  For this reason, we provide a detailed analytic description of the method in Appendix~A1.  The analytic arguments allowed us to check a number of intermediate steps in our method (Figures~\ref{theoryref} and~\ref{mockgenvtheory}), and to estimate the required sample size for implementing this technique (equation~\ref{SN}).

It is also straightforward to extend our technique to study the clustering of the absorbers (whose redshifts are known) with galaxies in the SDSS photometric sample (whose redshifts are not known) to study the MgII absorber-galaxy cross correlation function.  This is the subject of work in progress.  

\section*{Acknowledgments}
We would like to thank G. Prochter and J. Prochaska for providing 
the catalog of Mg II absorption systems analyzed in this work, 
which was supported in part by NSF Grant AST-0507501.  

Funding for the creation and distribution of the SDSS Archive has 
been provided by the Alfred P. Sloan Foundation, the Participating 
Institutions, the National Aeronautics and Space Administration,
the National Science Foundation, the U.S. Department of Energy, the 
Japanese Monbukagakusho, and the Max Planck Society.  The SDSS web
site is http://www.sdss/org/

The SDSS is managed by the Astrophysical Research Consortium (ARC)
for the Participating Institutions.  The Participating Institutions
are the University of Chicago, Fermilab, the Institute for 
Advanced Study, the Japan Participation group, Los Alamos National 
Laboratory, the Max-Planck-Institute for Astronomy (MPIA), the Max-
Planck-Institute for Astrophysics (MPA), New Mexico State University,
University of Pittsburgh, University of Portsmouth, Princeton 
University, the United States Naval Observatory, and the University
of Washington.


\appendix

\section{Mock catalog construction and tests}\label{mocks}

To ensure that the background subtraction method we use in our luminosity function estimation yields robust, accurate results, we test it on a mock catalog of galaxies. This mock catalog will need to incorporate a model of galaxy clustering, for it is the clustering of galaxies over scales of $\approx$ 100 kpc---10,000 kpc which ensures the viability of our technique. Because galaxies cluster, there will be an over density of them, compared to random, which surround a selected galaxy and share its redshift, in addition to unassociated field galaxies randomly projected into the field. This over density of galaxies is all that remains when we subtract away the random field galaxies.  Were we to ignore galaxy clustering in our mock catalogs, our method would yield nothing, for all galaxies in the field of a selected one would be random, unassociated ones, and would thus be eliminated from our sample upon background subtraction. 

Due to the nature of our technique, we actually need to produce two mock catalogs: one catalog of mock galaxies projected near simulated Mg II absorption line systems, and one of mock galaxies projected near simulated reference QSOs. While the same mock galaxy generating code will be used to construct both catalogs, the method by which simulated Mg II absorption line systems and simulated reference points are chosen must necessarily be different. In what follows, we first present an analytical overview of our technique to give the reader greater intuition for how it works. We next describe our mock galaxy generation code, then detail how the simulated Mg II absorption systems and references, respectively, are chosen. Finally, we explain how both catalogs of simulated galaxies are compiled and present the results of applying our background subtraction technique to them.

\subsection{Analytic calculation of background subtraction technique}\label{reftheory}

We begin our discussion with an analytic calculation which shows how our technique works.  

In a flux limited sample which covers some fraction $f_{\rm sky}$ of the sky, the observed number of objects with apparent magnitude $m$ is 
\begin{eqnarray}
 N(m) &=& f_{\rm sky} \int_{0}^{\infty} dz\,\frac{dV(z)}{dz} 
        \int dM\,\phi (M|z)\, \nonumber\\
   && \qquad \times\ 
         \delta_{\rm D}\Bigl(m=M+5\log \frac{d_{\rm L}(z)}{10\ {\rm pc}}+k(z)\Bigr) \nonumber\\ 
   &=& f_{\rm sky} \int_{0}^{\infty} dz\,\frac{dV(z)}{dz}\, \nonumber\\
   && \qquad \times\  \phi\Bigl(m - 5\log \frac{d_{\rm L}(z)}{10\ {\rm pc}} - k(z)\Big|z\Bigr),
\label{noflittlem}
\end{eqnarray}
where $\phi(M|z)$ is the luminosity function at $z$, $d_{\rm L}(z)$ is the luminosity distance to an object at $z$, and $k(z)$ is its $k-$correction.  
The surface density (number per unit area) of objects is 
\begin{equation}
 n(m) = \frac{N(m)}{4\pi\,f_{\rm sky}}.
\end{equation} 
If we assign all these objects the same redshift (and $k$-correction), then \eqref{noflittlem} will also describe the shape of the ``luminosity" distribution which results, except for a constant shift.  If we do this assignment for a number of different choices of redshift, the distribution of ``luminosities" will be given by simply shifting this shape for each redshift and summing up the result.  Thus,
\begin{eqnarray}
 N_{\rm ran}(M) = \int dz_{\rm abs}\, \frac{dN}{dz_{\rm abs}}\,
                    \omega(z_{\rm abs})\,
                    \int_{m_{\rm min}}^{m_{\rm max}} dm\,n(m)\nonumber\\
 \quad \times\ 
  \delta_{\rm D}\Bigl(M=m-\frac{d_{\rm L}(z_{\rm abs})}{10\ {\rm pc}}-k(z_{\rm abs})\Bigr).
\label{nofbigmnosub}
\end{eqnarray}
Here $dN/dz_{\rm abs}$ is the distribution of redshifts to be assigned to the objects (in this case the distribution of absorber redshifts), and we have allowed for the possibility that angular sizes $\omega$ of fields associated with redshift $z_{\rm abs}$ may depend on $z_{\rm abs}$. If we explicitly include in \eqref{nofbigmnosub} our expression \eqref{noflittlem} for the surface density of objects, we obtain
\begin{eqnarray}
 N_{\rm ran}(M) =  \int dz_{\rm abs}\,\frac{dN}{dz_{\rm abs}}\,
                     \frac{\omega(z_{\rm abs})}{4\pi}
                     \int_{z_{\rm min}(M)}^{z_{\rm max}(M)} dz\,\frac{dV}{dz}\,\nonumber\\
  \quad \times\ 
         \phi\Bigl(M - 5\log \frac{d_{\rm L}(z)}{d_{\rm L}(z_{\rm abs})} 
                     - k(z) + k(z_{\rm abs}) \Big|z\Bigr),
\label{nofbigmsub}
\end{eqnarray}
In essence, \eqref{nofbigmsub} describes the expected distribution of absolute magnitudes of objects which have random angular positions in a field but the same redshift distribution as the absorber catalog. This is precisely what we want our catalog of mock galaxies projected near simulated reference QSOs to contain. We can simplify this expression further if, as in the case of our mock catalog, the luminosity function does not evolve and there are no $k$-corrections: 
\begin{eqnarray}
 N_{\rm ran}(M) &=& 
     \int dz_{\rm abs}\, \frac{dN}{dz_{\rm abs}}\,
      \int_{z_{\rm min}(M)}^{z_{\rm max}(M)} dz\,\frac{dV(z)}{dz}\,\nonumber\\
 && \ \times\ 
        \phi\Bigl(M - 5\log \frac{d_{\rm L}(z)}{d_{\rm L}(z_{\rm abs})}\Bigr).
\label{nofbigmmc}
\end{eqnarray}
We shall show in \S~\ref{makemock} that \eqref{nofbigmmc} provides an excellent description of the absolute magnitude counts of simulated galaxies in our mock catalog counterpart to the reference population of \S~\ref{method}.  

The above considers the case in which fields are centered on a random point on the sky. If instead fields are centered on objects which are correlated with other objects in the field, there will be an additional contribution which comes from this spatial correlation $\xi$.  Namely, 
\begin{eqnarray}
 N_{\rm \xi}(M)  &\approx& 
 \int_{z_{\rm min}(M)}^{z_{\rm max}(M)} dz_{\rm abs}\, \frac{dN}{dz_{\rm abs}}
      \ \phi(M|z_{\rm abs}) \nonumber\\
 && \quad\times\ 2\pi \int_{r_{\rm min}}^{r_{\rm max}}
              dr_p\, r_p\int_{-\infty}^\infty dy\,\xi(r_p,y|z_{\rm abs}),
\label{corrnofbigm}
\end{eqnarray}
where we assume that luminosity distances $d_{\rm L}(z)$ and $k$-corrections do not change appreciably over the range of scales on which $\xi$ is not negligible, and that $\xi$ does not depend on luminosity.  It is this extra term which the background subtraction technique isolates.  

To gain intuition about this term, suppose that $\xi$ does not evolve over the redshift range spanned by the absorbers.  Then the term on the second line of \eqref{corrnofbigm} above is simply a constant (and, under the current hypothesis, independent of $M$).  For example, if the correlation function had the form $\xi(r) = (r_0/r)^2$, then 
\begin{equation}
 N_{\rm \xi}(M)  \approx V_\xi \, 
     \int_{z_{\rm min}(M)}^{z_{\rm max}(M)} dz_{\rm abs}\, 
     \frac{dN}{dz_{\rm abs}} \ \phi(M|z_{\rm abs}) ,
\label{xicorrnofbigm}
\end{equation}
where 
\begin{equation}
 V_\xi = 2\pi\, (\pi r_0^3)\, \frac{r_{\rm max} - r_{\rm min}}{r_0}.
\label{defineaxi}
\end{equation}
Further, if we assume the luminosity function does not evolve over the range of redshifts spanned by $z_{\rm abs}$, we could further reduce \eqref{xicorrnofbigm} to 
\begin{equation}
 N_{\rm \xi}(M) \approx V_\xi\, \phi(M)
 \int_{z_{\rm min}(M)}^{z_{\rm max}(M)} dz_{\rm abs}\, (dN/dz_{\rm abs}).
\label{ncorrofbigm}
\end{equation}
The right hand side of \eqref{ncorrofbigm} is proportional to the luminosity function times the number of fields lying in the redshift range wherein an object of absolute magnitude $M$ would have been observed in a flux limited catalog.  

Let $N_{\rm abs}$ denote the total number of absorbers in a catalog.  If we define 
\begin{equation}
 F_{\rm abs}(M) \equiv N_{\rm abs}^{-1}
 \int_{z_{\rm min}(M)}^{z_{\rm max}(M)} dz_{\rm abs}\, (dN/dz_{\rm abs}),
\end{equation}
then 
\begin{equation}
 N_{\rm \xi}(M)  \approx V_\xi \,\phi(M)\, N_{\rm abs}\,F_{\rm abs}(M).
 \label{NxiFabs}
\end{equation}
For $\bar n_{\rm abs}=0.001/(h^{-3}$Mpc$^3)$, $r_0=5h^{-1}$Mpc, $r_{\rm max} = 1h^{-1}$Mpc and $r_{\rm min} = 0.01h^{-1}$Mpc, we find that $N_{\rm \xi}(M) \approx 0.5\, \phi(M)\, N_{\rm abs}\,F_{\rm abs}(M)$.  

Notice that if the absorbers were uniformly distributed in comoving volume, i.e., 
 $dN/dz_{\rm abs} = \bar n_{\rm abs}\,f_{\rm sky}\,dV/dz_{\rm abs}$, 
then 
\begin{equation}
 N_{\rm \xi}(M) \approx \bar n_{\rm abs}\,V_\xi\,\phi(M)\,\,
                        f_{\rm sky}\,[V_{\rm max}(M) - V_{\rm min}(M)].
\label{nofmuniformdndz}
\end{equation}
If one then weights objects with luminosity $M$ by the inverse of $f_{\rm sky} \ [V_{\rm max}(M) - V_{\rm min}(M)]$, then the resulting distribution will be proportional to the luminosity function $\phi(M)$.  The constant of proportionality is the product of the number density of absorbers and the correlated volume.  In general, the absorbers will not be uniform in comoving volume, e.g., because of the QSO redshift distribution, or S/N issues with the spectrograph.  If so, the appropriate weighting factor is $F_{\rm abs}(M)$ rather than $V_{\rm max}(M)$.

\subsection{Mock catalog construction}\label{makemock}

Having provided an analytical description of our technique, we are ready to detail the construction of our mock catalogs, beginning with our mock galaxy generating code. Before we doing so, however, we must consider how in our construction to place galaxies near each other so as to mimic a clustering signal. Ideally, this would be done by populating dark matter halos produced in N-body simulations with galaxies according to the halo distribution function. Since we do not have such simulations to work with, we chose to mimic galaxy clustering by placing galaxies into groups of fixed comoving spherical volume, each group containing 20 galaxies. This is not meant to be an accurate description of real galaxy clustering properties, but rather a toy model through which we can illustrate the effects of clustering.  Within a group, galaxies are uniformly distributed within a spherical volume comoving radius $288h^{-1}$~kpc (i.e., the comoving volume of each group is $0.1h^{-3}$~Mpc$^3$). Projected onto the sky, the angular size of such a group at a redshift $z = 0.82$ is 30 arcseconds; at a redshift $z = 0.37$ it is 1 arcminute, and at a redshift $z = 0.1$ it is 3.4 arcminutes. (The background cosmology is $\Lambda$CDM with parameters given in the main text.)

Having chosen a method for placing galaxies into groups, we are ready to populate a simulated volume with them. A 1.56 square degree patch of sky spanning the redshift range $ 0 \leq z \leq 1$ forms the volume which we populate with mock galaxies. Galaxy groups are placed at random within this volume with number density $8.15\times 10^{-3} h^3 Mpc^{3}$ and their redshifts calculated accordingly. Each group is then assigned a random position within this patch of the sky. The angular size on the sky of each group is computed (this depends on its assigned redshift); its 20 member galaxies are then each assigned its redshift as well as a sky position. Luminosities for each galaxy are then drawn according to the \citet{blan03} luminosity function, evolved to the mean redshift ($z = 0.594$) of our Mg II absorption sample by the prescription of \citet{lin00}. This evolved luminosity function has $\phi^*=1.54 \times 10^{-2}$~($h^{-1}$Mpc)$^{-3}$, $M_{r}^*=-21.44$, and $\alpha=-1.04$. We shall hereafter refer to this luminosity function as the "input luminosity function." Once all galaxies have a redshift and absolute $r-$magnitude, their apparent $r-$magnitudes can be determined. 

Each mock galaxy now has a redshift, an angular position on the sky, an absolute magnitude $M_r$ and an apparent magnitude $m_r$; this is all the data we need to select either a simulated Mg II absorption system or a simulated reference QSO and find all galaxies projected near it.  We refer to this set of objects as the ``full" mock catalog.  A subset of these objects will satisfy the SDSS magnitude limit of $m_r=22.5$; we call this the ``apparent magnitude limited" mock catalog.
\begin{figure}
 \centering
 \includegraphics[width=\hsize]{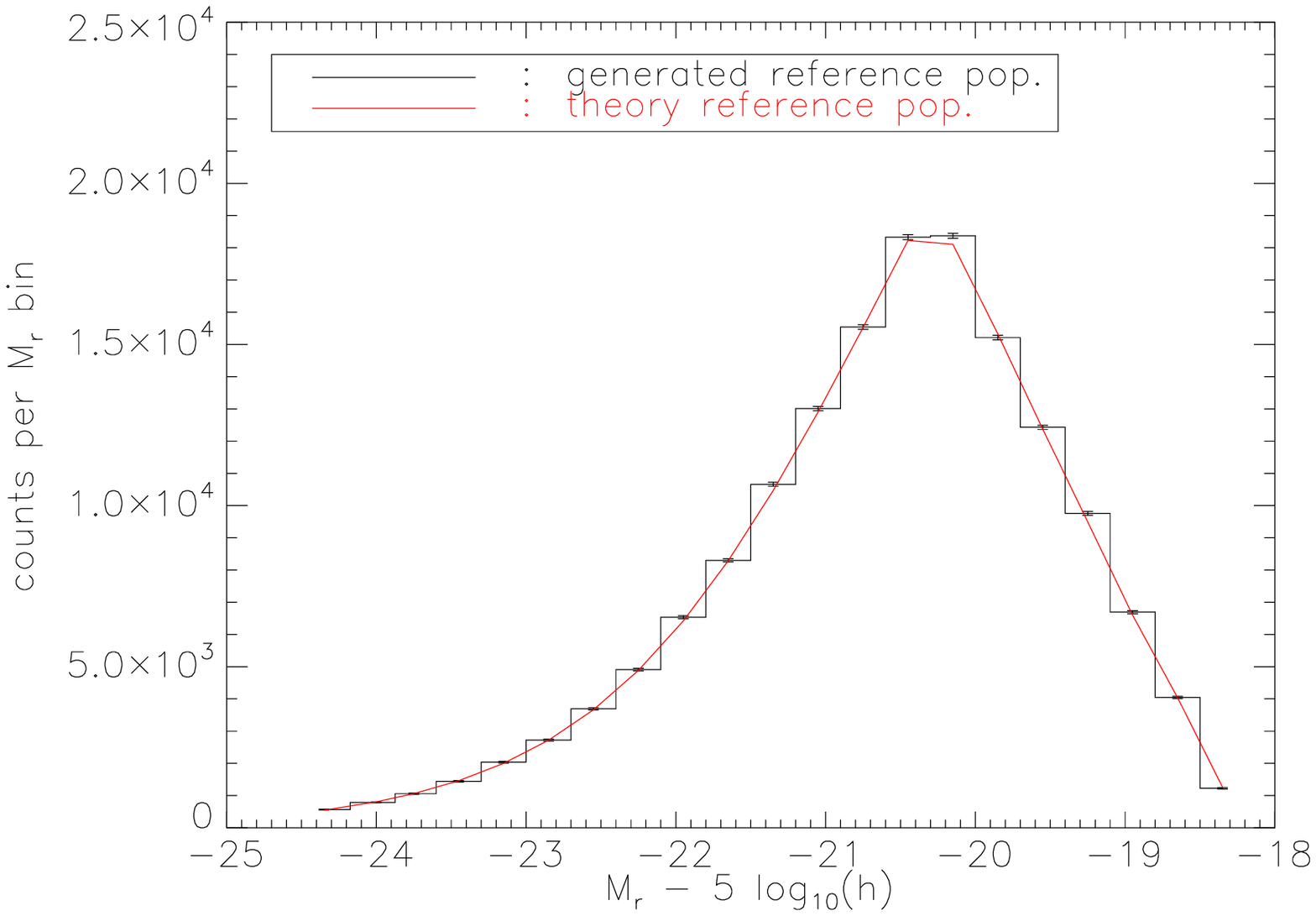} 
 \includegraphics[width=\hsize]{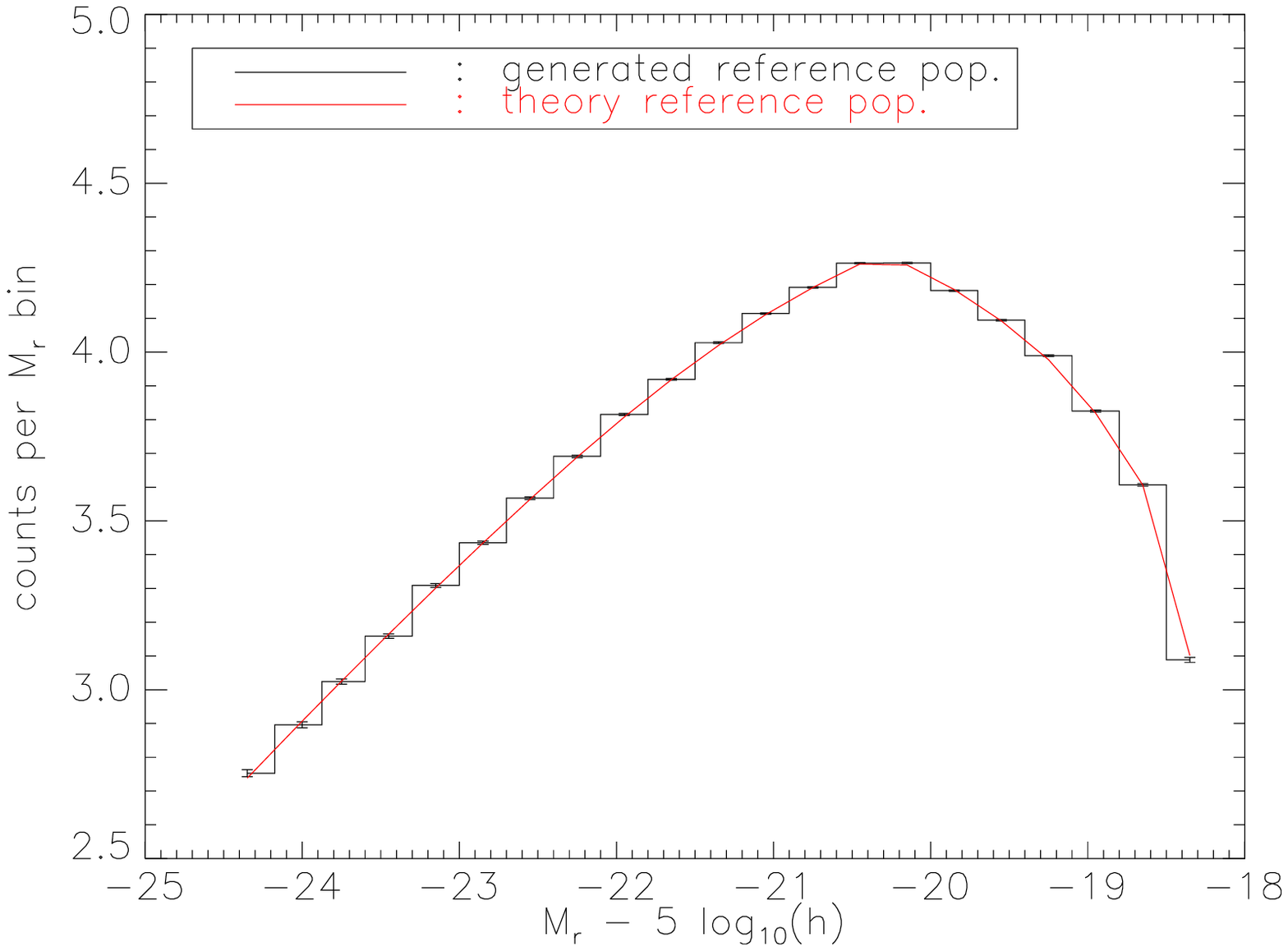}
 \caption{Absolute magnitude counts per bin obtained from the procedure outlined in \S ~\ref{reftheory} and the mock reference neighbour catalog. The black histogram shows the counts from the mock reference neighbour catalog; the smooth red curve is calculated as in \S ~\ref{reftheory}. The top panel displays the counts on a linear scale; in the bottom one, they are shown on a log scale.}
 \label{mockgenvtheory}
\end{figure}

We must now choose a galaxy from the ``full" mock catalog to serve as our mock Mg II absorber (selecting absorbers from the ``apparent magnitude limited" catalog is inappropriate, since detection of a galaxy by Mg II absorption is independent of its apparent brightness). We would like the set of such galaxies to match as closely as possible the redshift distribution of the Mg II absorption systems in our data. To achieve this, we randomly select an absorber from the SDSS data set and choose a mock absorber from our ``full" catalog that has the same redshift. We add one additional requirement, which is that the distance to the nearest object in the ``apparent magnitude limited" mock catalog match that of the nearest neighbour seen in the SDSS data, {\em and} that it has a similar apparent magnitude. Once our mock Mg II absorber has been chosen, we find those galaxies in the ``apparent magnitude limited" catalog located within 3 arcminutes of its position. Once found, they are re-assigned the redshift of the mock absorber and compiled into a final catalog, hereafter referred to as the ``mock Mg II neighbour catalog". We repeat the process of finding a mock Mg II absorber and its detectable neighbours--that is, those neighbouring galaxies with $m_{r} \leq 22.5$--until there is one mock Mg II system for each real one in our data sample, and all their neighbouring ``detectable" galaxies have been added to the mock Mg II neighbour catalog. 

The next step is to create a counterpart to the reference sample.  For each absorber with known redshift, we now select a random sky position to serve as a mock ``reference QSO". (A random sky position is chosen because our mock ``reference QSO" position need not correspond to a mock galaxy's, as it did for the mock Mg II systems.) We ensure that this random point does not coincide with that of a mock galaxy in the redshift range $0.37 < z < 0.82$ of the ``full" mock catalog, for in this case it could potentially have been a mock absorber.  We then search for mock galaxies in the ``apparent magnitude limited" subset which are located within 3 arcminutes of this position. These objects are all assigned the same redshift (that of the absorber), and then compiled into what we call the ``mock reference neighbour catalog".  We repeat this three times for each absorber in our data set.

To test that all has worked well, the histogram in Figure~\ref{mockgenvtheory} shows the counts per absolute magnitude bin in the mock reference neighbour catalog described above.  The solid red line shows our analytic calculation from the previous section (see \ref{nofbigmmc}). The correspondence between the two is excellent.  

\begin{figure}
 \centering
 \includegraphics[width=\hsize]{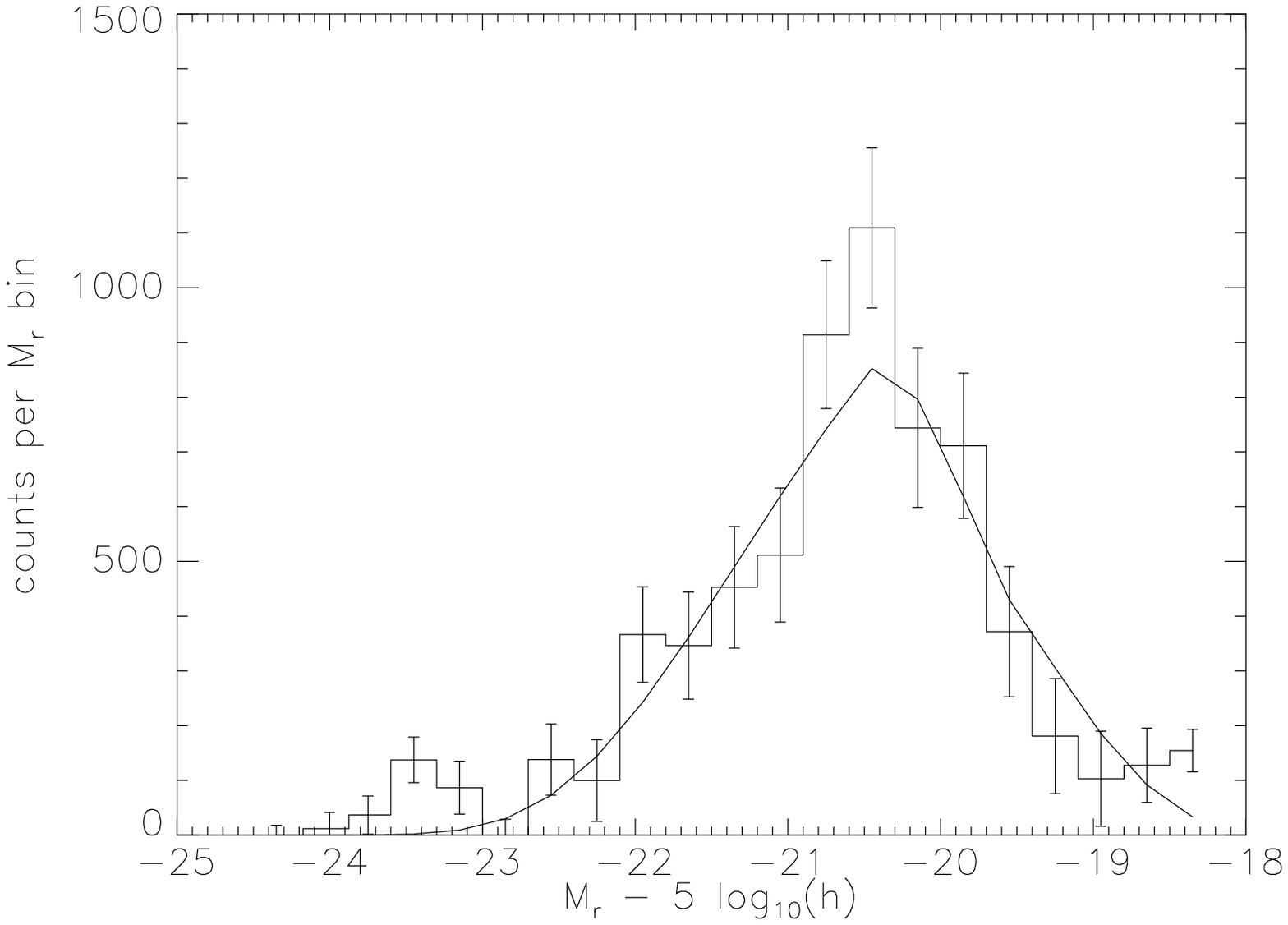}
 \caption{Absolute magnitude counts per bin obtained from the background subtraction of our mock Mg II neighbour and mock reference neighbour catalogs. The construction of these two mock catalogs is outlined in the text. The solid line is the expected shape of the distribution and is calculated as in \S ~\ref{mocks}.}
 \label{mockcountsperbin}
\end{figure}

With our mock Mg II neighbour and mock reference neighbour catalogs now compiled, we apply the procedure outlined in \S ~\ref{method} to estimate the absolute magnitude distribution of the galaxies they contain. The result of doing so is shown in Figure~\ref{mockcountsperbin}.  The solid line in this figure is our analytic calculation from equation \eqref{ncorrofbigm}). Though noisy, the shape of the absolute magnitude distribution mirrors the expected one. This demonstrates that our method can successfully recover the underlying absolute magnitude distribution of galaxies correlated with MgII absorption systems.

In the mock catalog, there are no $k$-corrections and the luminosity function does not evolve.  In this case, equation~\eqref{NxiFabs} suggests that the $N(M)$ distribution can be turned into an estimate of $\phi(M)$ simply by applying an appropriate $M$ dependent weight to each galaxy.  In practice, neither $k$-corrections nor evolution can be ignored, so this complicates the relation between $N(M)$ and $\phi(M|z)$.  This is why, in the main text, we present plots of $N(M)$ but not of $\phi$.  

\label{lastpage}
\end{document}